\documentclass[reprint,12pt]{JASA}

\begin{document}

\title[JASA/OMOQ for TSM]{An Objective Measure of Quality for Time-Scale Modification of Audio}


\author{Timothy Roberts}
\author{Kuldip K. Paliwal}
\affiliation{Signal Processing Laboratory, Griffith University, 170 Kessels Road, Nathan, QLD 4111, Australia}
\email{timothy.roberts@griffithuni.edu.au}

\preprint{Roberts, Griffith University}

\date{\today} 

\begin{abstract}
Objective evaluation of audio processed with Time-Scale Modification (TSM) remains an open problem.  Recently, a dataset of time-scaled audio with subjective quality labels was published and used to create an initial objective measure of quality. In this paper, an improved objective measure of quality for time-scaled audio is proposed.  The measure uses hand-crafted features and a fully connected network to predict subjective mean opinion scores. Basic and Advanced Perceptual Evaluation of Audio Quality features are used in addition to nine features specific to TSM artefacts.  Six methods of alignment are explored, with interpolation of the reference magnitude spectrum to the length of the test magnitude spectrum giving the best performance.  The proposed measure achieves a mean Root Mean Squared Error of 0.487 and a mean Pearson correlation of 0.865, equivalent to 98th and 82nd percentiles of subjective sessions respectively.  The proposed measure is used to evaluate time-scale modification algorithms, finding that Elastique gives the highest objective quality for Solo instrument and voice signals, while the Identity Phase-Locking Phase Vocoder gives the highest objective quality for music signals and the best overall quality.  The objective measure is available at https://www.github.com/zygurt/TSM.

   
\end{abstract}


\maketitle



\section{Introduction}
\label{sec:1}
Time-Scale Modification (TSM) is the process of modifying the duration of a signal without modifying the pitch of the signal.  In order to justify the quality of the processing, subjective testing must be undertaken.  However, it is expensive and time consuming.  Objective methods are available for evaluation of audio quality, however these methods require reference and test signals of identical duration.  Consequently, most published objective measures cannot be applied to this context.  Two objective measures, $SER$ by \citet{Verhelst_Roelands_1993} and $D_M$ by \citet{Laroche_Dolson_1999_IPL}, have been proposed.  However, they are shown to be only high level indicators of `phasiness' or quality \citep{Laroche_Dolson_1999_IPL}.  In this work, we propose the first effective objective measure of quality for time-scale modified audio.  It uses hand-crafted features with deep-learning methods and is trained using a recently published dataset \citep{SMOS_2020}.

Objective measures of quality seek to predict the quality of a test signal and can be broadly classified into two classes, traditional and machine learning.  Traditional measures such as Perceptual Evaluation of Speech Quality of \cite{ITU_P862_PESQ}, STOI of \cite{Gomez_2011_STOI} and the TSM specific measures of $SER$ and $D_M$ are purely analytical in nature.  Machine learning methods use neural networks to develop a relationship between subjective evaluations of the test signal and hand-crafted or data driven features extracted from reference and test signals, as in Perceptual Evaluation of Audio Quality (PEAQ) \citep{ITU_BS1387_PEAQ}.  Deep learning allows for objective measures that do not require a reference file, as in \citet{Avila_2019} for speech quality.  However this these methods have not yet been applied to TSM.

Training of deep learning methods requires a large amount of labelled signals.  Recently, a dataset of time-scaled audio with subjective labels was published for this purpose \citep{SMOS_2020}.  Reference files were drawn from a large variety of sources including speech, singing, solo harmonic and percussive instruments as well as a variety of musical genres.  The training subset, containing 5,280 processed files, was generated using six methods to time-scale 88 reference files at 10 ratios.  The methods used were the Phase Vocoder (PV) by \citet{Portnoff_1976}, the Identity Phase-Locking Phase Vocoder (IPL) by \citet{Laroche_Dolson_1999_IPL}, Waveform Similarity Overlap Add (WSOLA) by \citet{Verhelst_Roelands_1993}, Fuzzy Epoch Synchronous Overlap-Add (FESOLA) by \citet{Roberts_2019_FESOLA}, Harmonic Percussive Separation Time-Scale Modification (HPTSM) by \citet{Driedger_Muller_Ewert_2014} and Mel-Scale Sub-band Modelling (uTVS) by \citet{Sharma_2017}.  Playback speeds of 0.3838, 0.4427, 0.5383, 0.6524, 0.7821, 0.8258, 0.9961, 1.381, 1.667, and 1.924 were used as time-scale ratios ($\beta=\frac{1}{\alpha}$) for the training subset.  The testing subset, containing 240 files, was created using three additional methods to time-scale 20 reference files at a random $\beta$ in each band of $0.25<\beta<0.5$, $0.5<\beta<0.8$, $0.8<\beta<1$ and $1<\beta<2$.  Elastique by \citet{elastique}, the Phase Vocoder using fuzzy classification of bins (FuzzyPV) by \citet{Damskagg_2017} and Non-Negative Matrix Factorisation Time-Scale Modification (NMFTSM) by \citet{Roma_Green_Tremblay_2019} were used to generate the testing subset.  Finally, an evaluation subset was generated by processing the testing subset reference files with all previously mentioned methods, in addition to the Scaled Phase-Locking Phase Vocoder (SPL) by \citet{Laroche_Dolson_1999_IPL}, IPL and SPL variants of PhaVoRIT ($\overline{\textrm{IPL}}$ and $\overline{\textrm{SPL}}$) by \citet{Karrer_Lee_Borchers_2006} and Epoch Synchronous Overlap-Add (ESOLA) by \citet{Rudresh_2018}. 20 time-scale ratios in the interval of $0.22<\beta<2.2$ were used, resulting in 5,200 files with 400 files per method.  During subjective testing 42,529 ratings were collected from 263 participants in 633 sessions resulting in a minimum of 7 ratings per file. Subjective median opinion scores (MedianOS) and subjective mean opinion scores (SMOS) and  before and after normalization were provided as labels.  The dataset was published under the Creative Commons Attribution 4.0 International (CC BY 4.0) license at http://ieee-dataport.org/1987 \citep{SMOS_2020}.

The International Telecommunications Union (ITU) Recommendation BS.1387, more commonly known as Perceptual Evaluation of Audio Quality (PEAQ) \citep{ITU_BS1387_PEAQ}, is an objective measure of quality (OMOQ) developed primarily for evaluation of audio codecs.  It combines research from multiple groups and was released as an ITU standard in 2001.  PEAQ has two modes of operation, Basic and Advanced.  The Basic version (PEAQB) consists of an Fast Fourier Transform (FFT)-based peripheral ear model, pre-processing, calculation of 11 Model Output Variables (MOVs) and a small neural network.  The Advanced version (PEAQA) follows the same framework, but with a filter-bank-based ear model and 5 MOVs.

The FFT-based ear model aims to process the input signals in a similar way to the ear.  The model contains an FFT, rectification, scaling of the input signal, outer and middle ear weighting, auditory filter bands, internal noise, frequency-domain spreading and time-domain spreading.  The filter-bank model is identical in aim and contains scaling of the input signals, DC rejection, auditory filter band decomposition, outer and middle ear weighting, frequency-domain spreading, rectification, time-domain spreading, adding of internal noise and additional time-domain spreading.  Pre-processing of the resulting excitation patterns for both ear models creates patterns used in the calculation of the MOVs, the details of which can be found in \citet{ITU_BS1387_PEAQ}, \citet{Thiede_2000} and \citet{Kabal_2002}.

The basic MOVs can be categorised into six groups.  Modulation difference MOVs, \textit{WinModDiff1B}, \textit{AvgModDiff1B} and \textit{AvgModDiff2B}, are the windowed and linear averages of the modulation differences.  Noise loudness MOVs, of which \textit{RmsNoiseLoudB} is the only one used in the basic method, is the squared average of the noise loudness and takes masking into account.  Bandwidth MOVs, \textit{BandwidthRefB} and \textit{BandwidthTestB}, estimate the mean bandwidth of the reference and testing signals considering only frames with a bandwidth greater than 8kHz. Psuedocode for the calculation is given in \citet{ITU_BS1387_PEAQ}.  When considering auditory masking, \textit{Total NMRB}, is the linear mean of the noise-to-mask ratio, while Relative Disturbed Frames Basic, \textit{RelDistFramesB}, is the number of frames with a noise-to-mask ratio above 1.5dB as a ratio of the number of frames for the signal.  For detection probability, Maximum Filtered Probability of Detection (\textit{MFPDB}) models the smaller impact of distortions at the beginning of the file on quality assessment.  Average Distorted Block (\textit{ADBB}), uses the number of frames with a distortion detection probably above 0.5 and is calculated according to section 4.7.2 in \citet{ITU_BS1387_PEAQ}.  Finally, the Harmonic Structure of Error (\textit{EHSB}) MOV measures the harmonic structure of the error signal, as strong harmonic structure may be transferred to the error signal.  The advanced model adds an additional four MOVs, while also using \textit{EHSB}. \textit{RmsModDiffA}, \textit{RmsNoiseLoudAsymA} and \textit{AvgLinDistA} are all calculated from the filterbank ear model excitation patterns, while \textit{SegmentalNMRB} is calculated from the FFT model. For full details, see \citet{ITU_BS1387_PEAQ} and \citet{Kabal_2002}.

PEAQ makes use of a neural network to map the MOVs to a single Distortion Index (DI) value.  The network used with the basic model is a fully connected network with a single hidden layer of three nodes and sigmoid activation.  Features are scaled to between 0 and 1 before input to the network using 
\begin{equation}
\label{eq:PEAQ_Input_Scaling}
\hat{MOV[i]} = \frac{MOV[i]-a_{min}[i]}{a_{max}[i]-a_{min}[i]}
\end{equation}
where $a_{min}$ and $a_{max}$ are scaling factors.  Finally, the DI is mapped to the final Objective Difference Grade (ODG) minimizing the Root Mean Square Error (RMSE).


The initial PEAQ standard \citep{ITU_BS1387_PEAQ} was found to contain errors and omit vital information required for a proper implementation of the standard.  \citet{Kabal_2002} clarified these errors and omissions, and provided a MATLAB implementation of the standard. Available implementations include PQeval \citep{Kabal_2002}, gstpeaq \citep{gstpeaq}, EAQUAL \citep{eaqual}, peaqb-fast \citep{peaqb} and PEAQPython \citep{PEAQPython}.

The paper is organized as follows: Section~\ref{sec:Method} presents the proposed OMOQ method; Section~\ref{sec:Results} presents feature and network results as well as a comparison of TSM algorithms.  Availability, future research and conclusions are presented in Sections~\ref{sec:6}, \ref{sec:7} and \ref{sec:8} respectively.
 
\section{Method}
\label{sec:Method}
In this section, the proposed TSM objective measure is described. It uses a neural net to infer the SMOS score from hand-crafted features computed from audio processed by TSM. It includes PEAQ features, with modifications described in subsection~\ref{subsec:Method:1} and additional features specific to TSM artefacts described in subsection~\ref{subsec:Method:2}.  Feature preparation is described in subsection~\ref{subsec:Method:3} and the neural network is described in subsection~\ref{subsec:Method:4}.

\subsection{Changes to PEAQ}
\label{subsec:Method:1}
Changes were made to PEAQ MOV generation to allow for the use of time-scaled signals assuming that a constant time-scale ratio was applied while processing the signal.

Signal preparation begins by summing all input channels, before DC removal and normalization to the maximum absolute value.  The proposed method uses full scale as $\pm$1 rather than the 16-bit integers of PEAQ.  A single channel is used in the proposed method as multi-channel TSM is rarely considered \citep{Roberts_2018_Stereo}.  Consequently, a single channel used for detection probability calculations in \citet{ITU_BS1387_PEAQ} section 4.7.  The beginning and end of test and reference files are determined as the first and last time the sum of the absolute of four consecutive samples exceeds 0.0061, as per \citet{ITU_BS1387_PEAQ} section 5.2.4.4.  Signals are then truncated to these lengths.  This removes frames with low energy at the beginning and end of the signals during averaging calculations, and synchronises the time-scaling starting point.  

PEAQ assumes an input sample rate of 48kHz, however the proposed method calculates features based on the sample rate of the input signals.  In the processing of this dataset a sample rate of 44.1kHz is used.  As a result, the proposed method uses frequency values calculated from the given bin values of \citet{ITU_BS1387_PEAQ}.  In the calculation of \textit{BandwidthRefB} and \textit{BandwidthTestB}, frequencies are used rather than bin indices.  Noise floor is calculated above 21kHz with 8kHz used as the bandwidth cutoff for bin inclusion during averaging.  PEAQ and the proposed method both assume that bandwidth will be reduced due to processing.

The reference signal before and after spectral adaptation is used as input for \textit{AvgLinDistA} calculation.  However, the ITU specification is unclear as to the which filter envelope modulation ($Mod[k,n]$ in eq. 57) to use in equation 67.  The proposed implementation uses the reference modulation in calculation of $s_{ref}$ and $s_{test}$ for equation 66 of \citet{ITU_BS1387_PEAQ}.

The final change to the ITU standard in the proposed method is the calculation of  \textit{RelDistFramesB}.  The proposed method uses the interpretation of \citet{Kabal_2002} as `related to' meaning the fraction of frames exceeding 1.5 dB.

Six methods of alignment were investigated during development, time-instance framing anchored to the reference or test signal, and four methods of interpolating magnitude spectrum frequency bins along the time-axis.  Time-instance framing extracts frames from the reference and test signals at identical time-instances by scaling the frame locations by $\beta$, such that $S_R = u \beta S_T$ where $u$ is the frame number, $S_R$ is the reference signal shift in samples and $S_T$ is the test signal shift in samples.  In cases where $\beta$ is not known, the ratio between the lengths of the truncated input signals is used.  


While alignment through re-sampling either the reference or test signal to be the same duration is not suitable, due to resulting changes in pitch, it is possible to re-sample or interpolate low bandwidth representations of the signals, as shown by \citet{Sharma_2017}.  In the proposed method, interpolation for  basic PEAQ features is applied prior to the ear model using one of four targets: the longest signal, the shortest signal, the reference signal or the test signal.  For advanced PEAQ features, interpolation to the test signal is applied after application of the ear model.  There is no requirement for the time-scale to be known during calculation of features, only that the time-scale of the processed signal has been modified by a constant amount.  Through a simple thought experiment, we can observe that as we extend signals through interpolation the transient components of the signal will also be extended, while the same transients will not be extended through time-instance framing.  As such it is necessary to consider all, and combinations of, alignment methods.

\subsection{Additional Features}
\label{subsec:Method:2}

When calculating PEAQ Bandwidth features, asymmetric thresholds are used with +10dB used for \textit{BandwidthRefB} and +5dB used for \textit{BandwidthTestB}.  Test Bandwidth calculated with a +10dB threshold (\textit{BandwidthTestNew}) has been included as an additional feature.

The two published traditional OMOQ were included as features in the proposed method.  \citet{Roucos_Wilgus_1985} used the Signal to Error Ratio ($SER$), which is calculated by
\begin{equation}
\label{eq:SER}
SER = 10\log_{10}\frac{\sum^{\text{U}-1}_{u=0}{\sum^{\frac{N}{2}}_{k=0}{|X_T|^2}}}
                     {\sum^{\text{U}-1}_{u=0}{\sum^{\frac{N}{2}}_{k=0}{(|X_R|-|X_T|)^2}}}
\end{equation}
where $X$ is shorthand for $X(u,k)$, $u$ is the frame number, $k$ is the frequency bin, $\text{U}$ is the total number of frames, $N$ is the frame size, $X_R$ is the Short Time Fourier Transform (STFT) of the reference signal and $X_T$ is the STFT of the test signal. It is a measure of the difference between the magnitude spectra of the reference and test signals.  Practically, $SER$ is bounded to a maximum of 80 to avoid possible infinite results when processing identical files.  This empirical value was the maximum finite feature value for identical files.

\citet{Laroche_Dolson_1999_IPL} proposed an objective `phasiness' measure ($\text{D}_\text{M}$) by measuring the consistency of the synthesis reconstruction.  It is calculated by 
\begin{equation}
\label{eq:DM}
\text{D}_\text{M} = \frac{\sum^{\text{U}-1}_{u=0}{\sum^{\frac{N}{2}}_{k=0}{(|X_T|-|X_R|)^2}}}
                         {\sum^{\text{U}-1}_{u=0}{\sum^{\frac{N}{2}}_{k=0}{|X_R|^2}}}
\end{equation}
Neither of these measures have seen continued use with each measure noted to be only a high level indicator of signal `phasiness' \citep{Laroche_Dolson_1999_IPL}.


One cause of `phasiness' is phase unwrapping errors that occur when the time-scaling parameter ($\alpha$) is not an integer \citep{Laroche_Dolson_1999_IPL}.  In this work we propose a method for estimating the level of `phasiness' by considering the phase progression of reference and test signals.  The proposed `phasiness' features track phase progression through time for reference and test tracks, accounts for the change of time-scale and calculates the difference between the resulting unwrapped phase progression.  Weighting is applied to the phase difference, with unity and magnitude spectrum weighting applied in separate features within the proposed method.  These features are calculated in the following manner.  The phase spectra of the reference and test signals are calculated using the STFT and adjusted to be between 0 and $2\pi$ using
\begin{equation}
\label{eq:Phasiness_1}
\angle \hat{X} = 
    \begin{cases}
        \angle X, \quad \quad \angle X>0\\
        \angle X+2\pi, \quad otherwise
    \end{cases}
\end{equation}
forming $\angle \hat{X}$.  $2\pi$ is then successively added to each bin until it is greater than the same frequency bin in the previous frame using
\begin{equation}
\label{eq:Phasiness_2}
\acute{X} = \min(\angle \hat{X} +2\pi P) > \angle \hat{X}(u-1,k)
\end{equation}
where $P \in \mathbb{Z}$.  The longer $\acute{X}$ is then interpolated to match the length of the shorter signal, forming $\widetilde{X}$.  The weighted angle difference ($\Delta \varphi$) can then be calculated using 
\begin{equation}
\label{eq:Phasiness_3}
\Delta \varphi = 
    \begin{cases}
    W(k) \cdot (\acute{X}_R-\beta \widetilde{X}_T), \quad \text{U}_T>=\text{U}_R\\
    W(k) \cdot (\beta \widetilde{X}_R- \acute{X}_T), \quad otherwise
    \end{cases}
\end{equation}
where weighting is calculated with
\begin{equation}
\label{eq:Phasiness_4}
W(k) = 
    \begin{cases}
        \frac{|X_R|}{\text{max}|X_R|}, \quad \text{U}_T>=\text{U}_R\\
        \frac{|X_T|}{\text{max}|X_T|}, \quad otherwise
    \end{cases}
\end{equation}
or $W(k) = 1$ for no weighting.  Once the angle differences have been calculated, the mean `phasiness' features, using No Weighting (\textit{MPhNW}) and Magnitude Weighting (\textit{MPhMW}), are calculated by taking the means of the absolute weighted difference in time and frequency dimensions.  The standard deviation of the absolute weighted difference (\textit{SPhNW} and \textit{SPhMW}) are calculated by taking the standard deviation of the frequency mean of the absolute weighted difference.  A number of additional measures were explored including power spectrum weighting, Fletcher-Munson curve weighting and the mean first difference along the time dimension, however they were found to be poor measures or contribute little towards the training of the prediction network.



\begin{figure}[ht]
    \centering
    \includegraphics[trim= {25 35 35 45},clip,width=\linewidth]{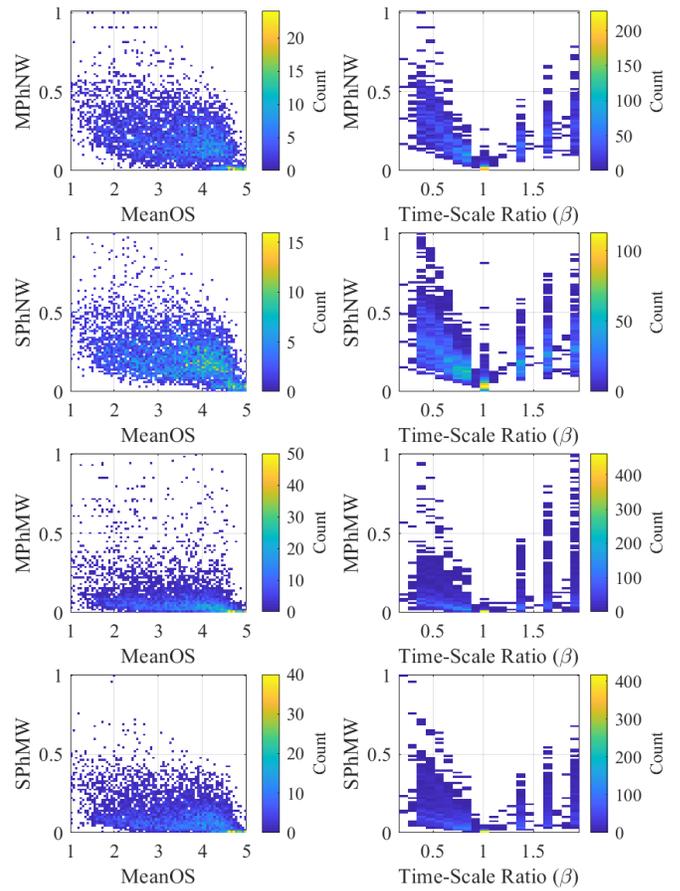}
    \caption{[Color Online] Phasiness features as functions of SMOS and TSM Ratio. Mean/Standard Deviation Phasiness No/Magnitude Weighting ([M/S]Ph[N/M]W).}
    \label{fig:Phasiness}
\end{figure}

Figure~\ref{fig:Phasiness} show the `phasiness' features compared to both SMOS and TSM ratio.  `Phasiness' can be seen to increase as the TSM ratio moves away from 100\% and as the SMOS decreases, as expected.  Animated 3-dimensional plots for all features color coded to each TSM method can be found in the supplementary material\footnote{\label{foot:Note1}See
Supplementary materials at zygurt.github.io/TSM/objective
for plots rotating between features as functions of SMOS and $\beta$.} and at zygurt.github.io/TSM/objective.

\begin{figure}[ht]
    \centering
    \includegraphics[trim= {25 0 35 20},clip,width=\linewidth]{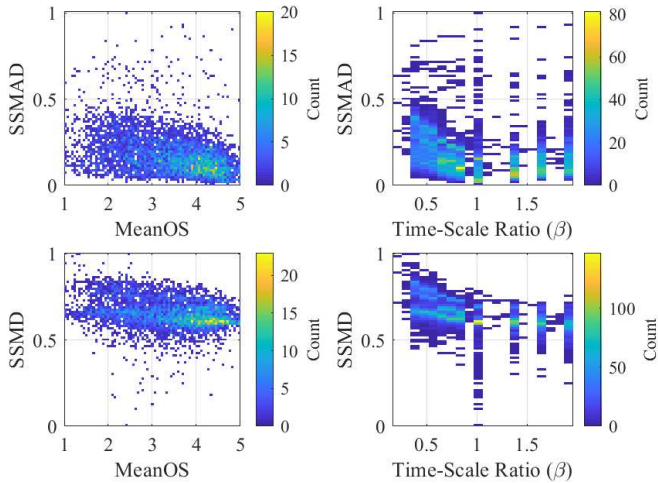}
    \caption{[Color Online] Spectral similarity features as functions of SMOS and TSM Ratio.}
    \label{fig:SS}
\end{figure}

`Phasiness' causes spectral coloration of the signal \citep{Laroche_Dolson_1999_IPL}, allowing for spectral similarity to be used as an indicator of phasiness.  Two features (\textit{SSMAD} and \textit{SSMD}) were developed using differences in the smoothed spectrum between reference and test signals.   Frames, aligned using reference frame anchors, are converted to normalized magnitude spectra using the STFT and Hann windowing.  Third-order polynomials are then fit to the spectra.  The resulting polynomials, without the intercept term, are applied to a linearly spaced vector $\frac{N}{2}$ in length.  Removal of the intercept term removes any overall level difference between the frames.  The mean absolute difference and mean difference between reference and test signals are calculated for each frame, with the means of these values forming the two spectral similarity features. These features also gives a measure of signal coloration introduced by the TSM algorithm.  Figure~\ref{fig:SS} shows the spectral similarity features in relation to the SMOS and TSM ratio.  Further analysis found groupings for individual and classes of TSM methods within the features. Time-domain methods inherently introduce less or no `phasiness', and FESOLA and WSOLA tend to have better spectral similarity than frequency-domain methods.  Refer to the \textit{SSMAD} and \textit{SSMD} animated graphs in the supplementary material\footnote{See footnote \ref{foot:Note1}} for examples. 

Changes in the transient content of the signal are common TSM artefacts.  Three features have been developed for the proposed method, Peak Delta, Transient Ratio and Harmonic Percussive Separation Transient Ratio, with no requirement for alignment between signals.  Peak Delta ($\Delta \text{P}$) is the difference in the number of onsets between the reference and test signals per second.  Onset detection is applied to both signals using the spectral features method described by \citet{Bello_2005}.  A weighting function, $W[k]=|k|$, is applied to the power spectrum
\begin{equation}
\label{eq:Onset1}
    \widetilde{E}[u] = \sum^{\frac{N}{2}-1}_{k=0}{W[k]|X|^2}
\end{equation}
before the first backward difference of the logarithmic transform is calculated using
\begin{equation}
\label{eq:Onset2}
    \Delta \widetilde{E}[u] = \text{log}_{10}\widetilde{E}[u]-\text{log}_{10}\widetilde{E}[u-1]
\end{equation}
Peak picking is applied to the onset results, where a peak is greater than its four surrounding values, with
\begin{equation}
\label{eq:Onset3}
    \text{P}[u] = 
        \begin{cases}
            1, \quad \Delta \widetilde{E}[u]>\Delta \widetilde{E}[u-2:u+2] \\
            0, \quad otherwise
        \end{cases}
\end{equation}
Finally, the difference in the number of peaks per second, calculated using
\begin{equation}
\label{eq:Onset4}
    \Delta \text{P} = \frac{f_s}{dim(x_R)}\left(\sum\text{P}_T[u] - \sum\text{P}_R[u]\right)
\end{equation}
is used as the feature, where $f_s$ is the sampling frequency and $dim(x_R)$ is the length of the reference signal in samples.

The transient ratio (\textit{TrRat}) is a measure of the change in transients due to processing and makes use of the peak locations calculated previously in equation \ref{eq:Onset3}.  It is calculated by selecting peaks where the onset peak level is greater than one standard deviation above the mean onset level using
\begin{equation}
\label{eq:Transient_Equation1}
    \hat{\text{P}} = \text{P}, \quad \text{where} \quad \Delta \widetilde{E}[\text{P}] > \overline{\Delta \widetilde{E}}+\sigma_{\Delta \widetilde{E}}
\end{equation}
Peak values are then used to calculate the ratio of mean transient level between the reference and test signals using
\begin{equation}
\label{eq:Transient_Equation2}
    \text{TrRat} = \frac{\frac{1}{\text{V}_R}\sum_{v=0}^{\text{V}_R-1}\Delta \widetilde{E}_R[\text{P}[v]]}
                        {\frac{1}{\text{V}_T}\sum_{v=0}^{\text{V}_T-1}\Delta \widetilde{E}_T[\text{P}[v]]}
\end{equation}
where $\text{V}$ is the total number of selected peaks, and $v$ is the index of the selected peaks.

\begin{figure}[ht]
    \centering
    \includegraphics[trim= {25 20 35 32},clip,width=\linewidth]{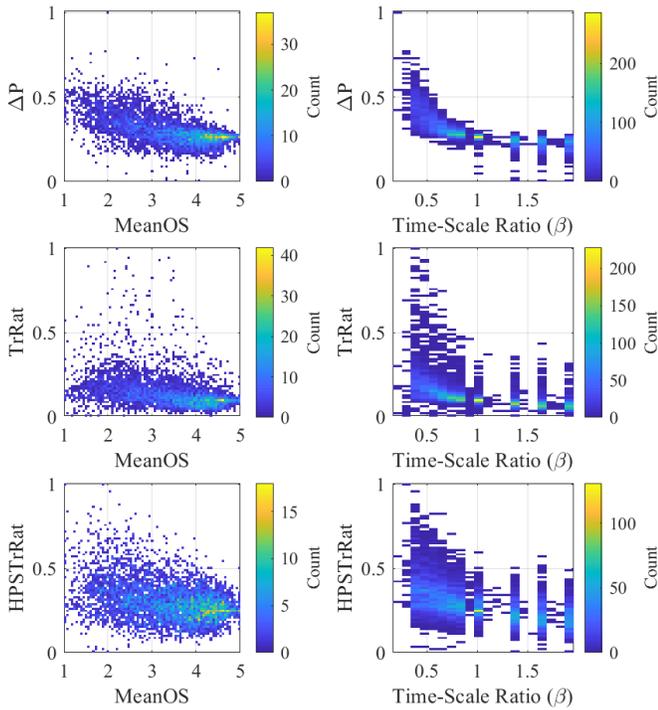}
    \caption{[Color Online] Transient features as functions of SMOS and TSM Ratio.}
    \label{fig:Transient}
\end{figure}

The Harmonic Percussive Separation Transient Ratio (\textit{HPSTrRat}) compares the Root Mean Square (RMS) levels of reference and test transients.  Transients are extracted from reference and test signals using the median filtering method of \citet{Driedger_Muller_Ewert_2014}.  The RMS of the extracted signals are calculated before the final feature is computed by the ratio of reference to test.   Figure~\ref{fig:Transient} compares each of the transient features to SMOS and TSM ratio.  

As musical noise is a known artefact introduced by TSM, it was also explored as a possible feature.  Spectral Kurtosis, as proposed by \citet{Torcoli_2019}, was explored using all previously discussed methods of alignment.  Lower, middle and upper frequency bands were used in addition to the maximum across all bands.  As all time-alignment methods produced highly correlated results, interpolation to test was chosen as the alignment method.  However, inclusion of these features reduced neural network performance and as a result they were removed from the features used in the final proposed network.

\subsection{Feature Preparation}
\label{subsec:Method:3}
Prior to network training, features were normalized for faster network convergence.  Each feature was scaled to the interval [0,1] using equation \ref{eq:PEAQ_Input_Scaling} where $a_{min}$ and $a_{max}$ are the minimum and maximum values for each feature. Target scores were also scaled to the interval [0,1] using 
\begin{equation}
\label{eq:Norm3}
    SMOS = \frac{SMOS-1}{4}
\end{equation}

\subsection{Network Structure}
\label{subsec:Method:4}
\begin{figure}[ht]
    \centering
    \includegraphics[width=1\linewidth]{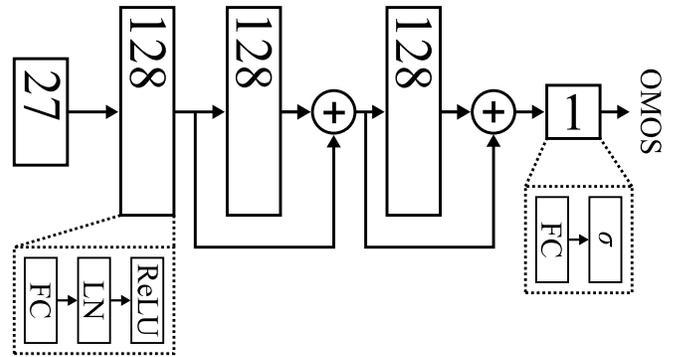}
    \caption{Neural network of proposed measure. Numbers denote layer nodes, FC is a Fully Connected layer, LN is Layer Normalization, ReLU activation function and $\sigma$ denotes a sigmoid activation layer.}
    \label{fig:FCN_NN}
\end{figure}

Estimation of opinion scores was formulated as a regression problem using a fully-connected neural network with three hidden layers of 128 nodes, shown in figure~\ref{fig:FCN_NN}.  Layer normalization and ReLU activation were used with residual connections around the second and third layers.  Residual connections are facilitated by adding the input of a layer to its output.  Sigmoid activation is applied to the final output.  

10\% of the training dataset was reserved for validation.  The network was trained for 800 epochs using a single batch, RMSE loss ($\mathcal{L}$), AdamW optimization \citep{AdamW_2017} and a learning rate of $1e^{-4}$.  Networks that were still improving after 800 epochs were trained for an additional 800 epochs.  Internal loss values were calculated using estimates in the interval of [0,1], while reported loss values were calculated using estimates scaled back to the original interval of [1,5].  As prediction of opinion scores for novel TSM methods is the network aim, early stopping based on validation loss was not used.  The optimal epoch was chosen as the epoch with the minimum overall distance ($\mathcal{D}$), calculated by
\begin{equation}
\label{eq:Overall_Distance1}
    \mathcal{D} = \|[\hat{\rho}, \hat{\mathcal{L}}]\|_{_2}
\end{equation}
where $\hat{\rho}$ and $\hat{\mathcal{L}}$ are calculated by
\begin{equation}
\label{eq:Overall_Distance2}
    \hat{\rho} = \|[1-\overline{\bm{\rho}},\Delta\bm{\rho}]\|_{_2}
\end{equation}
\begin{equation}
\label{eq:Overall_Distance3}
    \hat{\mathcal{L}} = \|[\overline{\bm{\mathcal{L}}},\Delta\bm{\mathcal{L}}]\|_{_2}
\end{equation}
where $\bm{\rho}=[\rho_{tr}, \rho_{val}, \rho_{te}]$, 
$\bm{\mathcal{L}}=[\mathcal{L}_{tr}, \mathcal{L}_{val}, \mathcal{L}_{te}]$, 
$tr$, $val$ and $te$ denote training, validation and testing,
$\overline{\bm{\mathcal{L}}}$ is the mean of $\bm{\mathcal{L}}$, 
$\overline{\bm{\rho}}$ is the mean of $\bm{\rho}$, 
$\Delta\bm{\rho} = max(\bm{\rho})-min(\bm{\rho})$ and 
$\Delta\bm{\mathcal{L}} = max(\bm{\mathcal{L}})-min(\bm{\mathcal{L}})$.  This allowed for the novel artefacts of the testing subset to inform the chosen optimal network, without their use in training the network.



\section{Results}
\label{sec:Results}
\subsection{Feature Results}
\label{subsec:Results:1}

\begin{figure}[ht]
    \centering
    \includegraphics*[width=\linewidth]{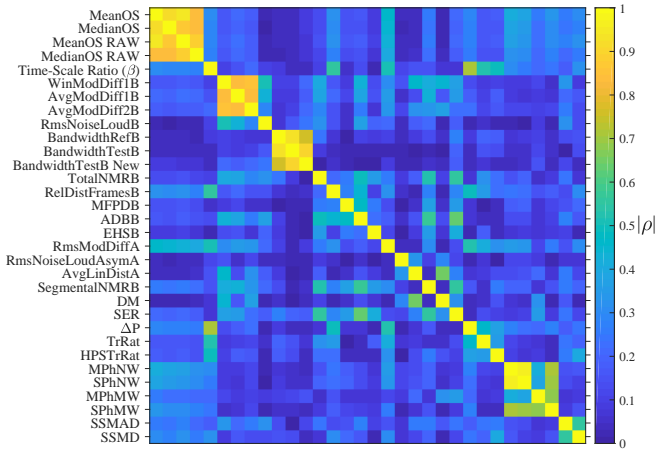}
    \caption{[Color Online] Feature correlation matrix for final features.  Average absolute correlation for $\beta<1$ and $\beta>1$ shown due to non-monotonic nature of relationship between features and time-scale ratio.}
    \label{fig:Feature_Corr}
\end{figure}

Features were manually pruned if the Pearson Correlation Coefficient (PCC) between features of the same type was above approximately 0.95, with figure~\ref{fig:Feature_Corr} showing the correlation between each of the features in the proposed measure.  This pruning increased the performance of the trained network.  Due to the non-linear nature of the relationship between $\beta$ and SMOS, correlation was calculated separately for $\beta<1$ and $\beta>1$ before averaging.  The additional features were found to have a greater correlation to the SMOS than most PEAQ features.  Of interest is the lack of individual features highly correlated with the SMOS or $\beta$, while still resulting in excellent network performance.

\subsection{Network Performance}
\label{subsec:Results:2}
\begin{figure}[ht]
    \centering
    \includegraphics[width=\linewidth]{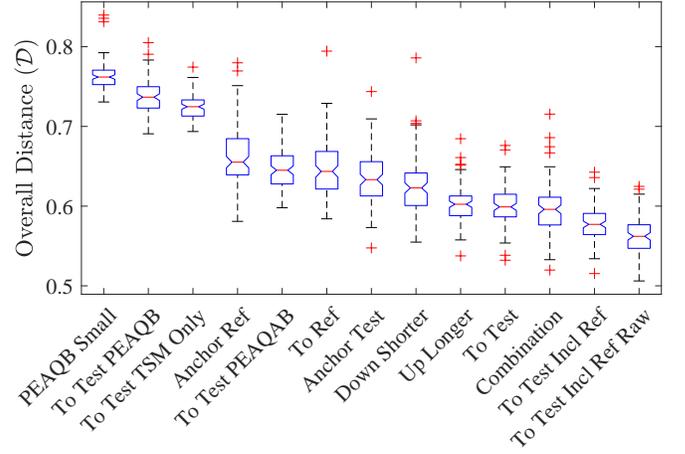}
    \caption{Box plot of best distance measure for each seed and training configuration ordered by median $\mathcal{D}$.  PEAQB NN uses original PEAQ network, all others use the network described in section \ref{subsec:Method:4}. Lower is better, less spread means less reliance on initial seed.}
    \label{fig:Overall_Distance_Boxplot}
\end{figure}

A wide range of testing and network configurations were considered during the development of the proposed method.  Network hyper-parameters were optimized through a systematic non-exhaustive search.  Each method of alignment was trained to SMOS, MedianOS, raw SMOS and raw MedianOS targets, where raw values were calculated prior to session normalization.  Additionally, baseline conditions, the inclusion of reference files within the training set, concatenation of logarithmic transforms of features and combinations of multiple alignment methods were considered.  Training of the network was conducted deterministically using seeds from 0 to 99. Figure~\ref{fig:Overall_Distance_Boxplot} shows the boxplot distribution of the best $\mathcal{D}$ for each of the seed values used while training to SMOS.  Lower values are better, with a smaller range meaning less reliance on the initial seed.

\begin{table*}[ht]
\caption{Mean RSME loss ($\overline{\bm{\mathcal{L}}}$) and range ($\Delta\bm{\mathcal{L}}$), mean PCC ($\overline{\bm{\rho}}$) and range ($\Delta\bm{\rho}$), median overall distance ($\widetilde{\bm{\mathcal{D}}}$) and minimum overall distance ($\text{min}(\bm{\mathcal{D}})$).  Best results in bold.}
\centering
\begin{ruledtabular}
\begin{tabular}{ccccccc}
\textbf{Features (Alignment)} & $\overline{\bm{\mathcal{L}}}$ & $\Delta\bm{\mathcal{L}}$ & $\overline{\bm{\rho}}$ & $\Delta\bm{\rho}$ & $\widetilde{\bm{\mathcal{D}}}$ & $\text{min}(\bm{\mathcal{D}})$ \\
\hline
Original PEAQB (To Test) & 0.668 & 0.054 & 0.719 & 0.075 & 0.762 & 0.731 \\
\hline
PEAQB (To Test) & 0.636 & 0.104 & 0.753 & 0.028 & 0.737 & 0.691 \\
\hline
TSM Only (To Test)  & 0.644 & 0.096 & 0.760 & \textbf{0.012} & 0.725 & 0.694 \\
\hline
All (Anchor Ref) & 0.529 & 0.169 & 0.842 & 0.062 & 0.655 & 0.581 \\
\hline
PEAQAB (To Test) & 0.558 & 0.109 & 0.820 & 0.043 & 0.645 & 0.598 \\
\hline
All (To Ref) & 0.537 & 0.149 & 0.835 & 0.053 & 0.644 & 0.584 \\
\hline
All (Anchor Test) & 0.505 & 0.145 & 0.854 & 0.052 & 0.633 & 0.548 \\
\hline
All (To Shorter) & 0.512 & 0.146 & 0.854 & 0.051 & 0.623 & 0.555 \\
\hline
All (To Longer) & 0.503 & 0.124 & 0.860 & 0.035 & 0.602 & 0.538 \\
\hline
All (To Test) & 0.500 & 0.108 & 0.856 & 0.031 & 0.599 & 0.532 \\
\hline
Combination (To Test and Anchor Test) & \textbf{0.458} & 0.200 & \textbf{0.882} & 0.080 & 0.596 & 0.520 \\
\hline
All (To Test Incl. Ref) & 0.487 & 0.096 & 0.865 & 0.029 & 0.577 & 0.515 \\
\hline
All (To Test Incl Ref Raw) & 0.478 & \textbf{0.071} & 0.853 & 0.029 & \textbf{0.562} & \textbf{0.506} \\
\end{tabular}
\end{ruledtabular}
\label{tab:RMSE_PCC}
\end{table*}

Across all test cases, the network was more successful when training to mean, rather than median, targets.  Consequently, the results discussed below will be solely focused on networks trained to mean targets.  To increase readability, median $\bm{\mathcal{D}}$ and best case $\bm{\mathcal{D}}$ with $\overline{\bm{\mathcal{L}}}$, $\Delta\bm{\mathcal{L}}$, $\overline{\bm{\rho}}$ and $\Delta\bm{\rho}$ values can be found in table~\ref{tab:RMSE_PCC}.  Values were calculated as per section \ref{subsec:Method:4}.

The baseline performance for the traditional methods was determined by correlation with the target. $SER$ and $D_M$ gave overall $\rho$ with subjective scores of 0.1445 and 0.0274 respectively. Machine learning baseline performance was obtained by applying time-aligned PEAQB features to the original PEAQB network described by \citet{ITU_BS1387_PEAQ}, shown as `Original PEAQB (To Test)'.  By increasing the complexity of the network, to that in section \ref{subsec:Method:4}, $\overline{\bm{\mathcal{L}}}$ and $\overline{\bm{\rho}}$ were improved, shown as `PEAQB (To Test)'.  Performance was further improved through the inclusion of PEAQ Advanced features, shown as `PEAQAB (To Test)'.  Interpolating to the length of the test signal was found to give the best performance followed by, in order, interpolating up to the longer signal, down to the shorter signal, anchoring frame locations to the test signal, interpolating to the reference signal length and anchoring frame locations to the reference signal. Using only the new TSM features gave similar performance to the PEAQ Basic features.  Concatenation of logarithmic transformations of features resulted in decreased network performance.  Including reference signals as test material, with targets set to 5, improved network performance and gave the best median overall distance for the seeds tested.  Combinations of features generated using interpolation to test and time-instance anchoring to test were also applied to the network.  This produced the best minimum distance after additional hyper-parameter tuning, but was highly reliant on initial seed selection and gave inconsistent results during TSM method evaluation.  Finally, combinations of concatenating logarithmic features, including reference signals and combining different alignment features were applied to the network, but all resulted in reduced performance.

Given the network performance in predicting raw SMOS outperforms prediction of normalized SMOS, investigation of Objective Mean Opinion Score (OMOS) differences was undertaken.  The mean difference between normalized and raw SMOS was found to be -0.0023, while the mean difference was found to be -0.004 for OMOS.   Normalizing was found to slightly extend the range of the SMOS values, with higher ratings for high quality files, and lower ratings for low quality files.  Given the \citet{ITU_BS1284} recommendation of normalization, and the improvement in all metrics after normalization \citep{Roberts_2020_SMOS_arXiv}, the final proposed objective measure of quality was trained to normalized SMOS using features aligned using interpolation to test, including reference files.



 

The proposed network achieved a best mean PCC of 0.865 and RMSE of 0.487.  These results place the proposed network at the 82nd and 98th percentiles of subjective sessions for PCC and RMSE respectively.
 
 
\subsection{TSM Algorithm Evaluation}
\label{subsec:Results:3}
TSM algorithms were compared using the evaluation subset of \citet{SMOS_2020}, described in section~\ref{sec:1}.  The uTVS implementation used in subjective testing ($\overline{\textrm{uTVS}}$), and an IPL by \citet{Driedger_Muller_2014}(DIPL) have also been included.  Although $\beta=1$ was used in the evaluation, in practice time-scaling is only applied at ratios other than 1.  Additionally, the minimum $\beta$ available for Elastique is 0.25.  Consequently, all results for $\beta=1$ and $\beta<0.25$ were excluded from averaging calculations.  Table~\ref{tab:Eval} shows the mean OMOS for each of the TSM methods tested in addition to means for each file class ordered by ascending overall mean.
\begin{table*}[ht]
\caption{Mean OMOS for each class of file and overall result. Means calculated without $\beta$ of 0.2257 and 1. Methods in order left to right are: NMFTSM, ESOLA, FESOLA, PV, FuzzyPV, Phavorit SPL, Subjective Testing uTVS, Phavorit IPL, uTVS, HPTSM, WSOLA, SPL, Elastique, Driedger's IPL and IPL.}
\centering
\begin{ruledtabular}
\begin{tabular}{cccccccccccccccc}
 & \textbf{NMF} & \textbf{ES} & \textbf{FES} & \textbf{PV} & \textbf{FPV} & $\overline{\textbf{SPL}}$ & $\overline{\textbf{uTVS}}$ & $\overline{\textbf{IPL}}$ & \textbf{uTVS} & \textbf{HP} & \textbf{WS} & \textbf{SPL} & \textbf{EL} & \textbf{DIPL} & \textbf{IPL} \\
\hline
Music & 2.842 & 2.883 & 2.982 & 3.571 & 3.641 & 3.643 & 3.548 & 3.673 & 3.622 & 3.627 & 3.516 & 3.660 & 3.691 & 3.750 & \textbf{3.810} \\
\hline
Solo & 2.826 & 3.604 & 3.621 & 3.416 & 3.408 & 3.540 & 3.623 & 3.595 & 3.650 & 3.581 & 3.754 & 3.618 & 3.773 & \textbf{3.833} & 3.742 \\
\hline
Voice & 2.876 & 3.244 & 3.341 & 2.968 & 3.179 & 3.137 & 3.235 & 3.125 & 3.270 & 3.350 & 3.441 & 3.415 & \textbf{3.617} & 3.533 & 3.568 \\
\hline
Overall & 2.848 & 3.207 & 3.282 & 3.344 & 3.433 & 3.460 & 3.477 & 3.485 & 3.525 & 3.530 & 3.565 & 3.574 & 3.694 & 3.710 & \textbf{3.717} \\
\end{tabular}
\end{ruledtabular}
\label{tab:Eval}
\end{table*}

Analysis is split into each class of reference file followed by overall average results.  In all cases, the noisy nature of the results for testing TSM method in \citet{Roberts_2020_SMOS_arXiv} has been smoothed. The poor performance of the uTVS subjective testing implementation, for $\beta$ close to 1, is also visible, with the updated implementation showing monotonic improvement towards $\beta=1$.

\begin{figure}
\fig{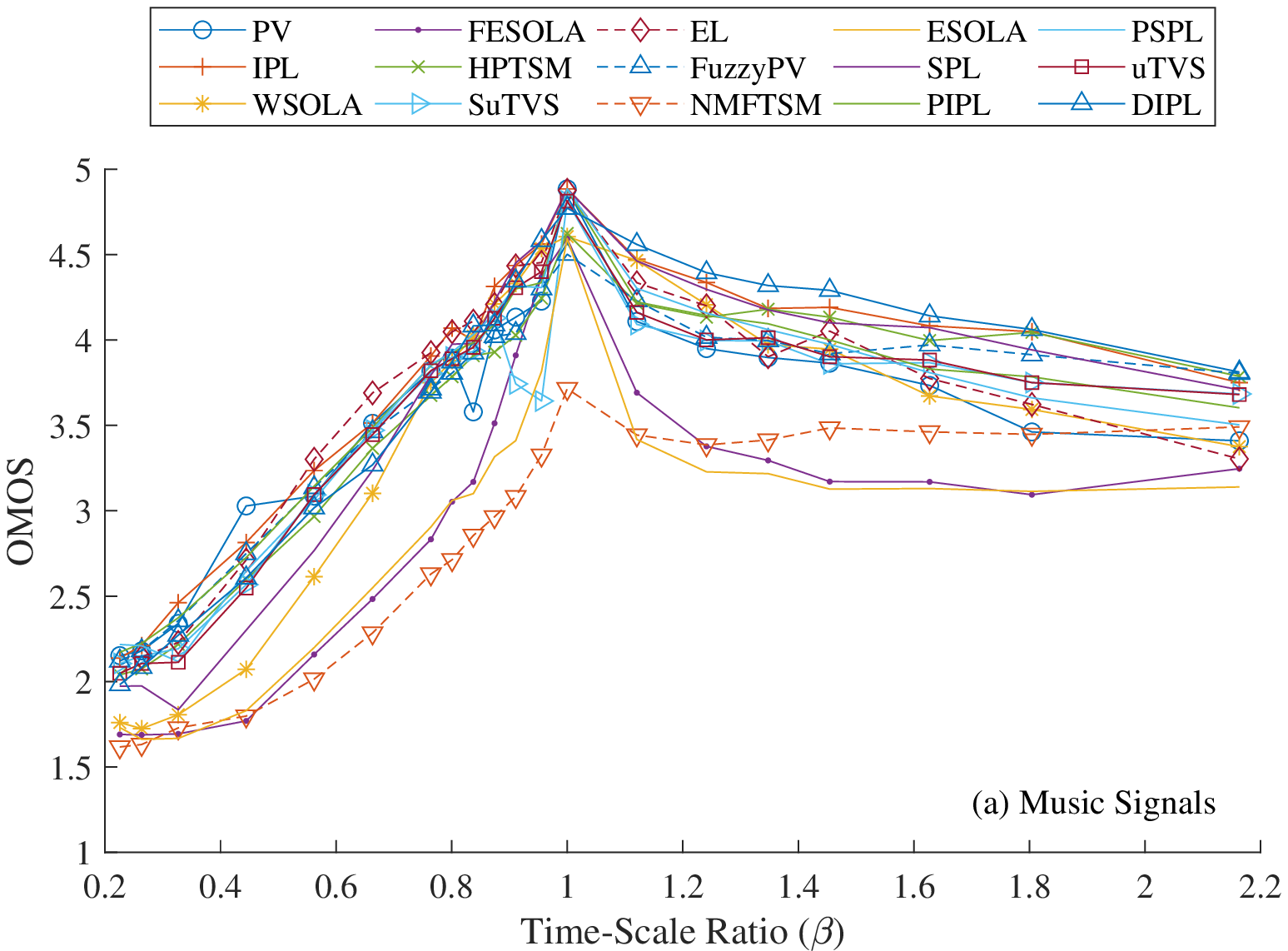}{2.92 in}{}
\fig{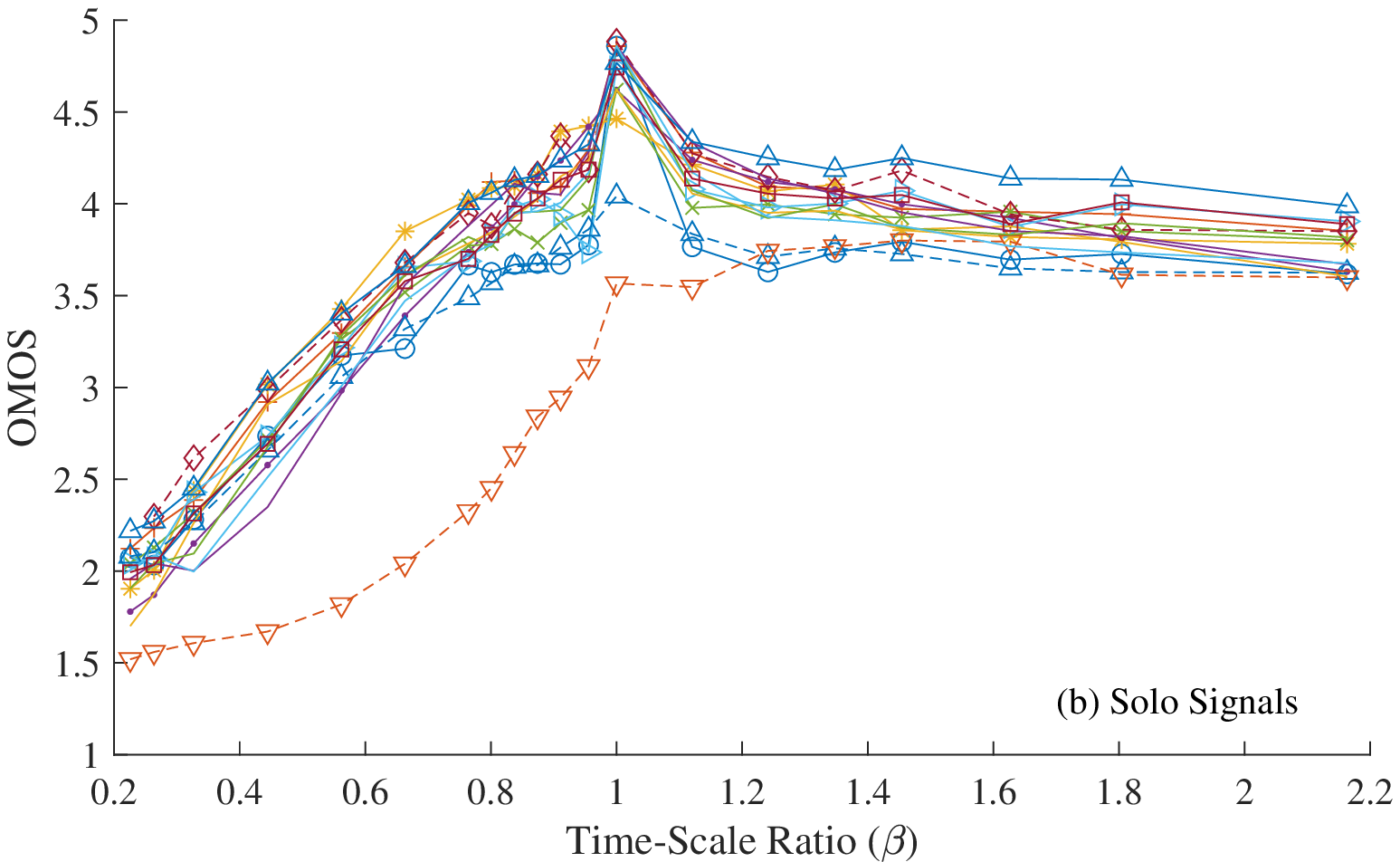}{2.92 in}{}
\fig{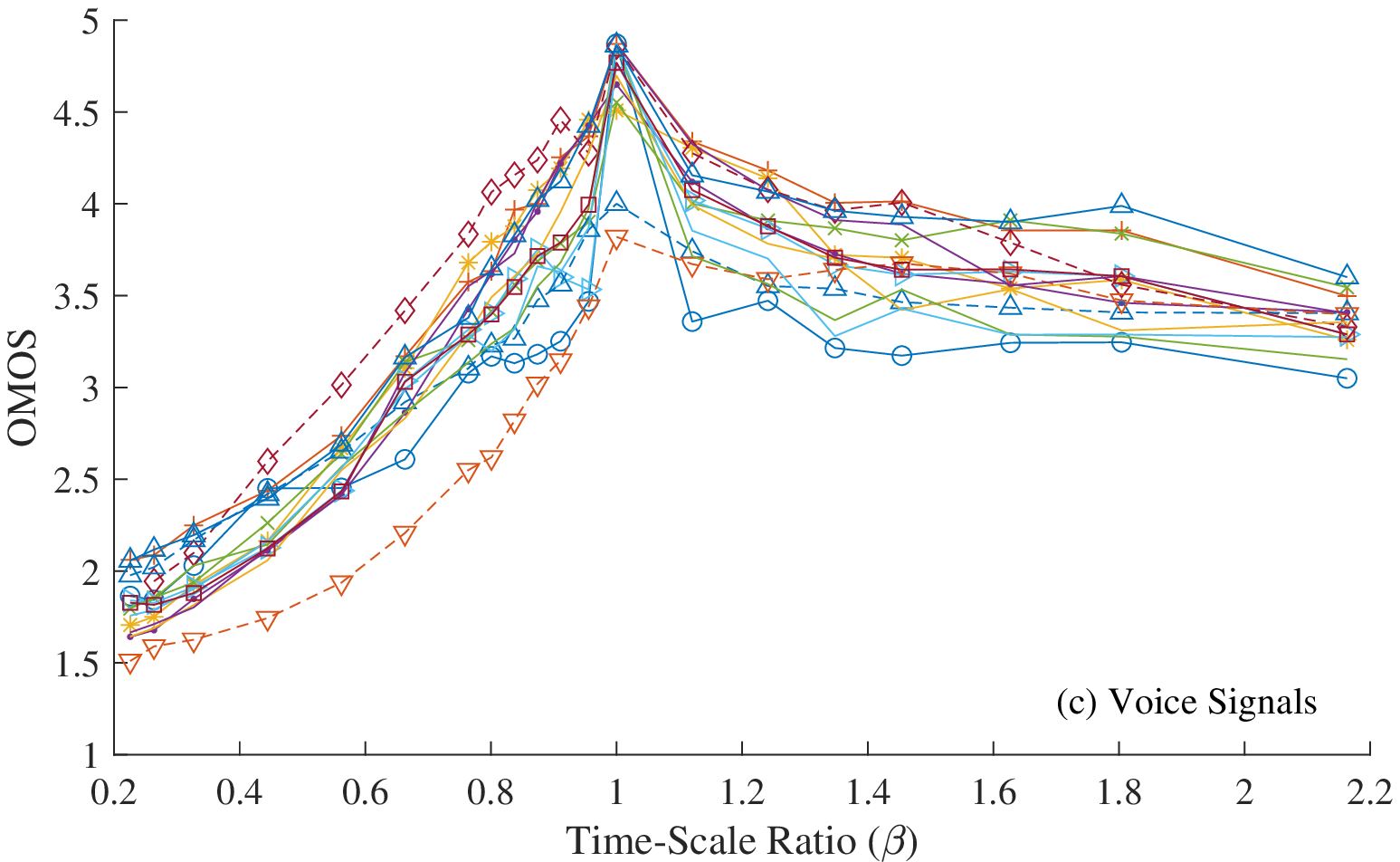}{2.92 in}{}
\fig{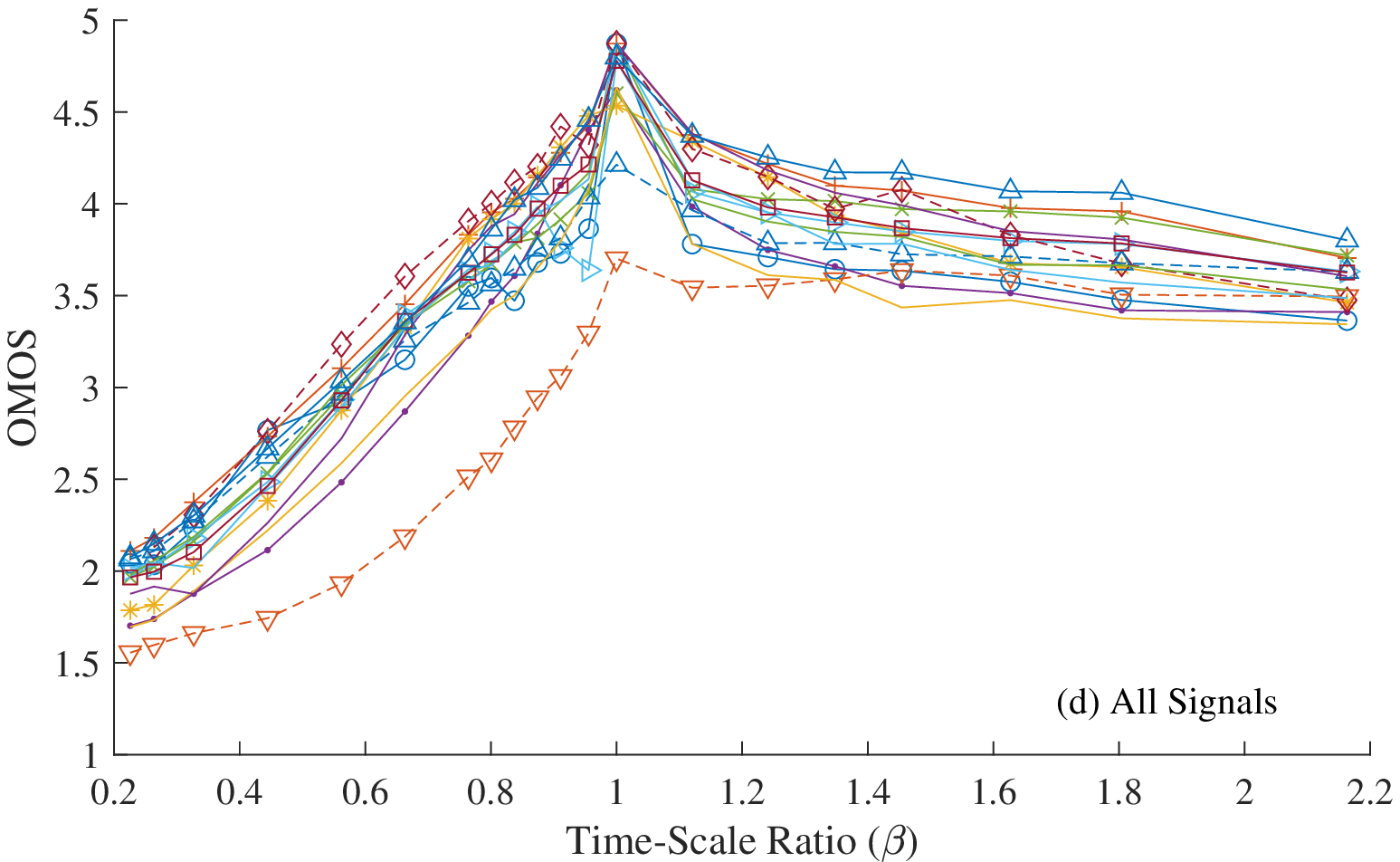}{2.92 in}{}
\caption{ \label{fig:Line_Graphs} [Color Online] Mean OMOS for each TSM method as a function of $\beta$ for: (a) Musical signals, (b) Solo signals, (c) Voice signals and (d) All signals combined.}
\end{figure}

For musical files, the OMOQ effectively differentiates between frequency and time-domain methods, where the quality worsens faster for time-domain methods.  WSOLA fairs the best of time-domain methods, diverging from frequency domain methods for $\beta<0.8$, as shown in figure~\ref{fig:Line_Graphs}(a).    When averaged, the OMOQ rates IPL highest followed by Elastique.  All other frequency domain methods gave similar results.


For solo files all methods except NMFTSM perform similarly with a maximum difference between methods of approximately 0.5 around $\beta=0.9$.  Method means at each time-scale can be seen in figure~\ref{fig:Line_Graphs}(b).  Driedger's IPL has the highest mean OMOS, followed by Elastique, WSOLA and IPL as shown in table~\ref{tab:Eval}.  The strong performance of WSOLA is  expected, due to individual harmonic and percussive signals.


Voice file OMOS shows the greatest variance between methods.  Of interest is the exponential shape of the curve for $\beta<1$ compared to the logarithmic shape for musical and solo classes, indicating harsher subjective evaluation of voice files was learned by the network.  Method means at each time-scale can be seen in figure~\ref{fig:Line_Graphs}(c).  Elastique has the highest mean OMOS, followed by IPL and WSOLA.  ESOLA and FESOLA give significantly improved performance for this class.

By averaging all OMOS, IPL has the highest average rating followed by DIPL and Elastique. Only 0.049 separates the following methods of SPL, WSOLA, HPTSM and uTVS.  The overall low performance of FuzzyPV is unexpected, given that it builds on IPL.  However other methods that perform decomposition of the signal, such as NMFTSM and HPTSM, also perform below the methods they build upon.  The overall means can be seen in figure~\ref{fig:Line_Graphs}(d). 


\section{Availability}
\label{sec:6}
The proposed tool is available from github.com/zygurt/TSM.  This includes the MATLAB scripts for feature generation, PyTorch code for the trained network and features for all dataset files in `csv' and `.mat' formats.  A bash script is also included that creates a virtual environment and installs required modules.  The features are also available with the subjective dataset at http://ieee-dataport.org/1987.

\section{Future Research}
\label{sec:7}
Future research is multi-faceted.  Evaluation of a wide range of commercial and lesser known published TSM methods should be considered in addition to comparisons of different implementations of the same TSM method.  Secondly, expansion into alternative and deeper neural networks should also be considered.  Initial testing resulted in a $\rho_{te}$ of 0.71 for a random forest network using the hand-crafted features, while using blind data-driven features created by a CNN used as input to a fully connected network resulted in a $\rho_{te}$ of 0.65.

\section{Conclusion}
\label{sec:8}
An objective measure for time-scaled audio was proposed with performance superior to most subjective listeners.  The measure used hand-crafted features and a fully connected network to predict subjective mean opinion scores.  PEAQ Basic and Advanced features were used in addition to nine novel features specific to TSM artefacts.  Six methods of alignment were explored, with interpolation of the magnitude spectrum to the length of the test signal giving the best performance.  The proposed measure achieved a mean RMSE of 0.487 and a mean PCC of 0.865.  Using the proposed method to evaluate algorithms, it was found that Elastique gave the highest objective quality for solo and voice signals, while the Identity Phase-Locking Phase Vocoder gave the highest objective quality for music signals and the best overall performance.  Future work includes optimisation of feature generation, exploration of other network structures and evaluation of more TSM algorithms.

\bibliography{OMOQ.bib}

\end{document}